\documentstyle[twoside,fleqn,espcrc2]{article}

\makeatother

\font \math=msym10 scaled \magstep 1
\newcommand{\mmath}[1]{{\mbox{\math #1}}}

\title
{Scattering in a Simple 2-d Lattice Model\thanks{Presented by C.B. Lang.
Supported by Fonds zur F\"or\-der\-ung der
Wissenschaftlichen Forschung in \"Oster\-reich, project P7849.} }
\author
{C.R. Gattringer\address{Max Planck Institut f\"ur Physik,
\\ F\"ohringer Ring 6, D-8000 Munich, Germany}, I. Hip and C.B. Lang
\address{Institut f\"ur Theoretische Physik,\\
Universit\"at Graz, A-8010 Graz, Austria}}

\begin{document}

\begin{abstract}
L\"uscher has suggested a method to determine phase shifts from the
finite volume dependence of the two-particle energy spectrum. We apply
this to two models in d=2: (a) the Ising model, (b) a system of two Ising
fields with different mass and coupled through a 3-point term, both
considered in the symmetric phase. The Monte Carlo simulation makes use
of the cluster updating and reduced variance operator techniques. For
the Ising system we study in particular O($a^2$) effects in the phase
shift of the 2-particle scattering process.
\end{abstract}

\maketitle

\section{INTRODUCTION}

Most of the particle entries in the annual review of particle
properties of the particle data group refer to resonances. Usually they
are created in two-particle collisions and observed in corresponding
cross sections and scattering phase shifts. However, in quantum field
theory unstable states still are one of the tantalizing problems.
In the lattice approach, up to now efforts have concentrated on the
identification of stable particle states. The determination of
phase shifts and resonance parameters has been reserved for a future of
more powerful computers and there are only very few (courageous)
contributions \cite{Lu86}- \cite{GaLa92}.

Recently, L\"uscher \cite{Lu91} has refined earlier ideas
\cite{Lu86,Wi89} on how to determine phase shifts in an elastic
two-particle system from the energy spectrum in a finite volume and
together with Wolff \cite{LuWo90} the power of the method was
demonstrated in the d=2 O(3) model. Here we discuss some of our results
\cite{GaLa92} where we use this method to determine the phase shifts in
a simple d=2 model with two particle types and resonating phase shifts.
Other contributions to this conference present first
results for d=3 \cite{Fi92} and d=4 \cite{Zi92} (cf. also
\cite{Ni92}) where life is harder and statistics scarcer. The case d=2
provides an excellent testbed to study some of the juicy details
of the approach.

\subsection{Method}

Let us briefly review the idea.
Consider the scattering of two identical particles of mass $m$ in a box
of finite spatial extension $L$. The time extension is assumed to be
sufficiently large, not to contribute to finite size effects.
The size of the spatial volume $L$ and the periodic b.c., however, are
responsible for the quantization of the momenta. In the elastic regime
$2m \leq W < 4 m$ (or $3 m$, depending on the theory) the allowed
momentum values $k$ are, in the two-dimensional case, related to the
scattering phase shift via the quantization condition
\begin{equation}\label{Phaseshiftrelation}
2 \delta(k_n) + k_n L = 2 n \pi \; ,\; n \in \mmath{N}  .
\end{equation}
Assuming vanishing total momentum (CMS) the total energy of the
2-particle state is just twice the energy of the back-to-back single
particle states
\begin{equation} \label{Wdispersionrelation}
W_n = 2 \sqrt{m^2+k_n^2}  .
\end{equation}
Thus given $m$ and a couple of low lying energy levels $W_n$ one
may obtain values of the (infinite volume) phase shift at
the corresponding values $k_n$. Varying $L$ one may cover a whole range
of momentum values. Relation (\ref{Phaseshiftrelation}) holds in that
simple form for d=2 but can be generalized to higher
dimensions \cite{Lu91}.

One has to take care of the following restrictions.
\begin{itemize}
\item The interaction region and the single particle correlation
length ought to be smaller than the spatial volume.
\item Polarization effects due to virtual particles running around
the torus should be under control.
\item Lattice artifacts will turn up in $O(a^2)$ corrections.
\item For the determination of the energy spectrum one should
consider correlation functions of sufficiently many observables.
\end{itemize}
In d=2 all these can be controlled.

\subsection{Model and Simulation}

We choose a model, where two light particles
$\varphi$ couple to a heavier particle $\eta$ giving rise to
resonating behaviour. The action is given by
\begin{eqnarray}\label{action}
S &=& -\kappa_\varphi \sum_{x \in \Lambda, \mu=1,2} \varphi_x
\varphi_{x+\hat{\mu}} \nonumber \\
  & & -\kappa_\eta \sum_{x \in \Lambda, \mu=1,2} \eta_x
\eta_{x+\hat{\mu}} \nonumber \\
  & & + \frac{g}{2}\sum_{x \in \Lambda, \mu=1,2} \eta_x \varphi_x
            (\varphi_{x-\hat{\mu}} + \varphi_{x+\hat{\mu}})    .
\end{eqnarray}
The values of the fields are restricted to $\{ +1, -1\}$. The sums run
over all sites $(x_0, x_1)$ of the euclidean $L\times T$ lattice
$\Lambda$ with periodic boundary conditions. The 3-point term was
introduced in a nonlocal but symmetric way, because $ \varphi_x^2 \equiv
1$. For $g=0$ this is just a system of two independent Ising models,
each describing in the scaling region interacting bosonic
fields with mass $m(\kappa)$.
The corresponding masses have been adjusted to $m_\varphi \simeq 0.19$
and $m_\eta \simeq 0.5$ ($m_\eta$ defined by the resonance peak position)
by calibrating the couplings $\kappa_\varphi$ and $\kappa_\eta$.

When kinematically allowed, the term proportional to $g$ gives rise to
transitions like $\eta \rightarrow \varphi \varphi$ rendering $\eta$ a
resonance in the $\varphi \varphi$ channel. We study the model at $g$ =
0, 0.02 and 0.04. Throughout this work we use $T=100$; the spatial
extension $L$ varies between $12$ and $60$. For each set of couplings
and lattice size we performed typically $2\times 10^5$ measurements.
Our Monte Carlo simulation utilizes the cluster updating method
introduced for the Ising model in \cite{SwWa87}. The statistical errors
are estimated with the Jackknife method. Details of the simulation
technique and the phase diagram can be found
in \cite{GaLa92}.

\section{OBSERVABLES}
\subsection{Single particle state}

The operator of a $\varphi$ state with momentum
$p_{1,\nu} = 2 \pi \nu /L, \nu = -L/2+1, \ldots, L/2 $,
is given through
\begin{equation}\label{Operatorphi}
\frac{1}{L}\sum_{x_1\in \Lambda_{x_0}}
\varphi_{x_0,x_1} \exp{(i x_1 p_{1,\nu})}  ,
\end{equation}
where $\Lambda_{x_0}$ denotes a timeslice of $\Lambda$. Its
connected correlation function over temporal distance $t$ decays
exponentially $\propto \exp{(-p_{0,\nu} t)}$ defining $p_{0,\nu}$;
in particular $p_{0,\nu=0} = m_\varphi$.

For the determination of the energy spectrum a precise knowledge of the
single particle mass and related finite size effects is important.
We find \cite{GaLa92}, that our results for $p_{0,\nu}$ follow with high
precision the
spectral relation for the lattice propagator of a Gaussian particle with
mass $m$,
\begin{equation} \label{latticerelation}
p_{0,\nu}= \cosh^{-1}( 1-  \cos p_{1,\nu}+ \cosh m ) .
\end{equation}
This expression deviates from the continuum dispersion relation (d.r.)
$p_{0,\nu} = (m^2 + p_{1,\nu}^2)^{1/2}$ by a leading correction
$O( (ap)^2)$.

The observed mass, as compared to the ``real'' mass at vanishing lattice
spacing and infinite volume, is also affected by polarization due to
self interaction around the torus. We confirm this behaviour and find
good agreement with the expected exponential decrease \cite{Lu89}.
We also determined the wave function renormalization
constants for the fields \cite{GaLa92}.

\begin{figure}[t]
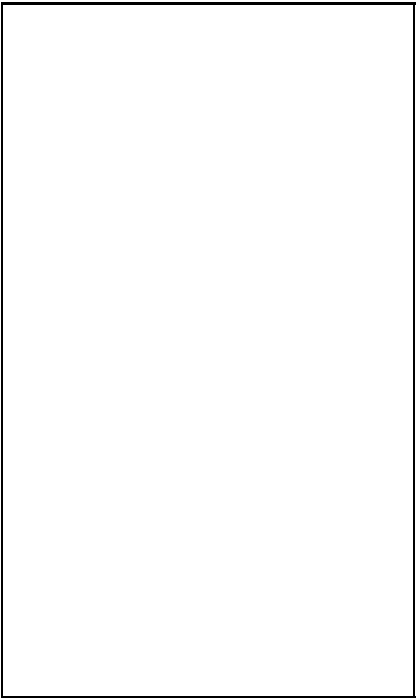

\framebox[55mm]{\rule[-21mm]{0mm}{90mm}}
\caption{Phase shifts for the Ising model for two values of the mass,
determined with the continuum d.r. (2) ;
full line: $\delta =-\pi/2$, dashed line: 2-particle threshold.}
\label{fig1}
\end{figure}
\begin{figure}[t]
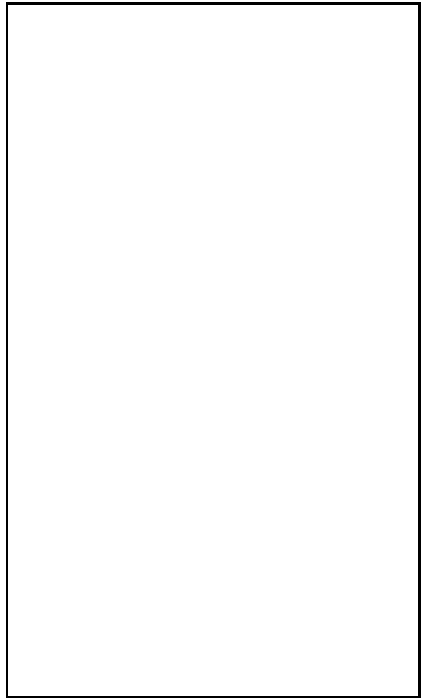

\framebox[55mm]{\rule[-21mm]{0mm}{90mm}}
\caption{Like fig. 1, but now using the lattice d.r. (9) }
\label{fig2}
\end{figure}

\subsection{Scattering sector}

We consider operators with total zero momentum and quantum numbers of
the $\eta$,
\begin{eqnarray}
\label{Operators}
N_1(x_0) &=&\frac{1}{L}\sum_{x_1\in \Lambda_{x_0}} \eta_{x_1,x_0}, \\
N_j(x_0) &=&\frac{1}{L^2}\sum_{x_1,y_1\in \Lambda_{x_0}}
e^{ip_j(x_1-y_1)} \varphi_{x_1,x_0} \varphi_{y_1,x_0}
,\nonumber \\
&&\mbox{with~}p_j=\frac{2\pi(j-2)}{L} ,\; j=2,3\ldots
\end{eqnarray}
and measure all (connected) cross-correlations
$M_{nm}(t) = \langle N_n(t) N_m(0)\rangle_c$
($t$ denotes the separation of the time slices).
The operators $N_{j>1}$ describe two $\varphi$-particles in the
CM system with relative momentum $2 p_j$.
Because of the interaction they do not correspond to eigenstates
of our model. Indeed they are eigenstates of the Gaussian model of free
bosons. However, if the set is complete, a diagonalization of
the correlation matrix provides the necessary information on the energy
spectrum of this channel.
The transfer matrix formalism yields the spectral decomposition
\begin{equation} \label{Spectraldeco}
M_{n m} ( t ) = \sum_{l=1}^{\infty} {v^{(l)}_n}^* v^{(l)}_m e^{- t W_l} ,
\end{equation}
where $v^{(l)}_n = \langle l | N_n 0 \rangle$
are the projections of the states $| N_n 0 \rangle$ (generated by the
operators $N_n$ out of the vacuum) on the energy eigenstates
$\langle l|$ of the scattering problem.

The number of operators considered should be chosen larger than the
number of states in the elastic regime $2 m_\varphi \leq W < 4
m_\varphi$ and not larger than $L/2$ to be linearly independent. A
larger set provides a better representation of the eigenstates but
enhances the numerical noise.

We work with between 4 and 6 operators
depending on $L$ and distances $t = 1\ldots 8$.
Details on the determination of the eigenspectrum in the
scattering sector and on the representation of the eigenstates
are discussed in \cite{GaLa92}.

\section{RESULTS FOR THE PHASE SHIFTS}
\subsection{Resonance}

In \cite{GaLa92} we present our results for the
observed energy levels and the resulting phase shifts. Since the pure
Ising model has an S-matrix equal to $-1$ \cite{SaMiJi77}, the phase
shift starts with $-\pi/2$ and then shows, for non-vanishing $g$, a
clear resonance behaviour manifesting itself in a fast increase by
$\pi$. It can be nicely approximated by a standard effective range
resonance formula \cite{Lu91,GaLa92}.

\subsection{$O(a^2)$ corrections}

Let us consider the Ising case ($g=0$) in more detail.
Here we increased the statistics significantly (1.5 million
measurements) and repeated the calculation for $m=0.19$ and a higher
value of the mass $m= 0.5$ (fig.\ref{fig1}). As mentioned we expect a
phase shift of $-\pi/2$ and the results are in agreement with this.
At small $k$ the energy is close to the 2-particle threshold and the
statistical error of the energy transforms via
(\ref{Wdispersionrelation}) into a relatively larger error of $k$ and
thus $\delta(k)$. Higher $k$, on the other hand, stem from large values
of the energy with intrinsically larger statistical fluctuations.
However, due to the enhanced statistics we do identify a systematic
deviation from $-\pi/2$ increasing with $k$ and $m$. We attribute this
behaviour to $O(a^2)$ corrections.

As mentioned earlier, the d.r. (\ref{Wdispersionrelation}) gives the
total energy of the
asymptotic 2-particle state which (under the assumption of localized
interaction region) is just twice the energy of the outgoing particles.
Now the $O(a^2)$ corrections of the single particle d.r. have
been nicely described by replacing the continuum d.r. by the
lattice relation (\ref{latticerelation}).
We therefore  replace (\ref{Wdispersionrelation}) by the corresponding
lattice expression,
\begin{equation} \label{newWdispersionrelation}
W_n = 2 \cosh^{-1}(1 -\cos k_n + \cosh m )    .
\end{equation}
Our data for $W_n$ now produce slightly different values of $k_n$ and
$\delta$ exhibited in fig.\ref{fig2}, in better agreement with a
constant value of
$-\pi/2$. We conclude that the leading $O(a^2)$ corrections can be
expressed by replacing the continuum d.r. by the lattice relation,
at least in the 2 dimensional Ising Model.
In general we suspect that lattice artifacts in the phase shift can be
diminished by studying carefully the dispersion relation of
single particle states.

\section{CONCLUSION}

We have determined phase shifts in the Ising model and resonating phase
shifts in a model with two types of particles and a three-point
coupling. We find that L\"uscher's suggestion for determining these
phase shifts is indeed a very reliable method, at least in d=2.
The leading $O(a^2)$ effects in the 1- and
2-particle channel may be explained by the differences between lattice
vs. continuum dispersion relations.


\end{document}